\input harvmac
\skip0=\baselineskip
\divide\skip0 by 2
\def\tmpsp{\the\skip0}

\def\skipthis#1{{}}

\def\p{\partial}

\Title{\vbox{\baselineskip12pt\hbox{hep-th/9911237}
\hbox{HUTP-99/A064}}}
{\vbox{\centerline{Mass Gap in Kaluza-Klein Spectrum}
	\vskip2pt\centerline{in a Network of Brane Worlds}}}
  
\centerline{Soonkeon Nam\foot{Permanent Address :
Dept. of Physics, Kyung Hee University; Seoul, 130-701, Korea, nam@string.kyunghee.ac.kr}
} 
\bigskip\centerline{Department of Physics}
\centerline{Harvard University}
\centerline{Cambridge, MA 02138}
\centerline{\it nam@pauli.harvard.edu}

\vskip .3in \centerline{\bf Abstract} 
We consider the Newton's force law for brane world consisting of
periodic configuration of branes. We show that it supports a massless
graviton. 
Furthermore, this massless mode is well separated from the Kaluza-Klein
spectrum by a mass gap. 
Thus most of the problems in phenomenology coming from continuum of Kaluza-Klein modes
without mass gap are potentially cured in such a model.

\smallskip
\Date{11/99}
\lref\rubakov{V. Rubakov and M.E. Shaposhnikov, Phys. Lett. {\bf 125B} (1983) 136.}
\lref\antoniadis{I. Antoniadis, Phys. Lett. {\bf 246B} (1990) 377.}
\lref\add{N. Arkani-Hamed, S, Dimopoulos, and G. Dvali, Phys.\ Lett.\ {\bf B429} (1998) 35.}
\lref\addk{N. Arkani-Hamed, S. Dimopoulos, G. Dvali, and N. Kaloper,
hep-th/9907209.}
\lref\aadd{I. Antoniadis, N. Arkani-Hamed, S. Dimopoulos, and G. Dvali, Phys.
Lett. {\bf 436B} (1998) 257.}
\lref\rs{L. Randall and R. Sundrum, Phys. Rev. Lett.
{\bf 83} (1999) 3370; hep-th/9906064.}
\lref\st{K. Skenderis and P.K. Townsend, hep-th/9909070. }
\lref\gog{M. Gogberashvili, hep-ph/9904383.} \lref\cs{C. Csaki and Y. Shirman,
hep-th/9908186.} \lref\nelson{A.E. Nelson, hep-th/9909001.}
\lref\dfgk{O. DeWolfe, D.Z. Freedman, S.S. Gubser, and A. Karch,
hep-th/9909134.}
\lref\bc{K. Behrndt and M. Cvetic, hep-th/9909058.}
\lref\bs{I. Bakas and K. Sfetsos, hep-th/9909041.}
\lref\bpsjunc{G.W. Gibbons and P.K. Townsend, Phys.\ Rev.\
Lett.\ , {\bf 83} (1999) 1727.}
\lref\bpsdom{K. Skenderis and P.K. Townsend, hep-th/9909070.}
\lref\Boucher{W. Boucher, Nucl.\ Phys.\ {\bf B242} (1984) 282.}
\lref\Townsend{P.K. Townsend, Phys.\ Lett.\ {\bf 148B} (1984) 55.}
\lref\lps{H. Lu, C.N. Pope, and E. Sezgin, Phys.\ Lett.\ {\bf B371} (1996)
46-50.}
\lref\cg{A. Chamblin and G.W. Gibbons, hep-th/9909130.}
\lref\saffin{P.M. Saffin, to appear, Phys. Rev. Lett., hep-th/9907066.}
\lref\mpp{E.W. Mirabelli, M. Perelstein, and M.E. Peskin, Phys. Rev. Lett. 
{\bf 82} (1999) 2236.}
\lref\mp{E.W. Mirabelli and M.E. Peskin, Phys. Rev. {\bf D58} (1998) 065002.}
\lref\cnamone{J. Casahorran and Soonkeon Nam, Int. J. Mod. Phys. {\bf A6} (1991) 5467.}
\lref\cgr{M. Cvetic, S. Griffies, and S. Rey, Nucl. Phys. {\bf B381} (1992) 301.}
\lref\rps{B.S. Ryden, W.H. Press, and D.N. Spergel, Ap. J.
{\bf 357} (1990) 293.}
\lref\network{Soonkeon Nam, hep-th/9911104.}
\lref\cht{S.M. Carroll, S. Hellerman, and M. Trodden, hep-th/9905217.}
\lref\oins{H. Oda, K. Ito, M. Naganuma, and N. Sakai, hep-th/9910095.}
\lref\susybreak{N. Arkani-Hamed and S. Dimopoulos, hep-ph/9811353.}
\lref\vilenkin{M. Aryal, A.E. Everett, A. Vilenkin, and  T. Vachaspati,
Phys. Rev. {\bf D34} (1986) 434.}
\lref\cv{S. Cecotti and C. Vafa, Commun. Math. Phys. {\bf 158} (1993) 569.}
\lref\manyfold{N. Arkani-Hamed, S. Dimopoulos, G. Dvali, and N. Kaloper,
hep-th/9911386.}
\lref\hstt{H. Hatanaka, M. Sakamoto, M. Tachibana, and K. Takenaga,
hep-th/9909076.}
\lref\brandhuber{A. Brandhuber and K. Sfetsos, hep-th/9908116.}
\lref\experiment{See for example T.G. Rizzo, hep-ph/9910255, and references
therein.} 
\lref\dunne{G. Dunne and J. Feinberg, Phys. Rev. {\bf D57} (1998) 1271.}
\lref\scarf{F.L. Scarf, Phys. Rev. {\bf 112} (1958) 1137.}
\lref\ashcroft{N.W. Ashcroft and N.D. Mermin, {\it Solid State Physics},  New
York, Holt, Rinehart and Winston, 1976.}
\lref\merzbacher{E. Merzbacher, {\it Quantum Mechanics} 2nd ed., New
York, John Wiley and Sons,  1970.}
\lref\cohentannouji{C. Cohen-Tannoudji, B. Diu, and F. Laloe, {\it Quantum
Mechanics},  New York, Wiley, 1992.}
\lref\lykken{J. Lykken and L. Randall, hep-th/9908076.}
\lref\generalization{W.D. Goldberger and M.B. Wise, hep-ph/9907218; hep-ph/9907447;
M.A. Luty and R. Sundrum, hep-th/9910202;
N. Arkani-Hamed, S. Dimopoulos, G. Dvali, and N. Kaloper, hep-th/9907209;
C. Csaki and Y. Shirman, hep-th/9908186;
I. Oda, hep-th/9908104;
A. Chodos and E. Poppitz, hep-th/9909199;
J. Cline, C. Grojean, and G. Servant hep-ph/9909496, hep-th/9910081;
P. Kraus, hep-th/9910149.}
\lref\supergravity{A. Kehagias, hep-th/9906204; H. Verlinde, hep-th/9960182.}
\lref\cosmology{P. Binetruy, C. Deffayet and D. Langlois, hep-th/9905012;
N. Kaloper, hep-th/9905210; T. Nihei, hep-ph/9905487; C. Csaki, M. Graesser,
C. Kolda, and J. Terning, hep-ph/9906513;
J.M. Cline, G. Grojean and G. Servant, hep-ph/9906523;
D.J. Chung and K. Freese, hep-ph/9906542; H.B. Kim and H.D. Kim,
hep-th/9909053;
P. Kanti, I.I. Kogan, K.A. Olive, and P. Pospelov, hep-ph/9909481;
C. Csaki, Michael Graesser, Lisa Randall, and John Terning, hep-ph/9911406.}
\lref\phenomenology{K.R. Dienes, E. Dudas, and T. Gherghetta, hep-ph/9908530;
H. Davoudiasl, J.L. Hewett,  and T.G. Rizzo, hep-ph/9909255.}
\lref\nanotube{R. Saito, G. Dresselhau, and M.S. Dresselhaus, {\it Physical
Properties of Carbon Nanotubes}, London, Imperial College Press, 1998.}
There has been a considerable interest in the model where the Standard Model
is confined to a (3+1) dimensional subspace in the higher
dimensions\add\aadd\rs. This is a renewal of old ideas\rubakov\antoniadis\ in
light of recent developments of string theory, especially due to the
fundamental role that extended objects, i.e. branes, play a fundamental
role.
There has been a lot of works related to this recently, such as generalization
of the RS model\lykken\generalization, in relation to supergravity or
superstrings\supergravity, in relation to cosmology\cosmology, and also some
phenomenological consequences\phenomenology.

An alternative understanding of
the hierachy problem of the electroweak and gravitational mass scales is one
of the major advantage of such a senario. In this senario, our understanding
of classical and quantum gravity has to be critically reanalyzed too: Newton's
law is affected by the Kaluza-Klein modes of the large extra dimensions, and
gravity may become strong at a scale of few TeV in the full $4+n$ dimensional
space. Furthermore, the effect of virtual exchange of Kaluza-Klein towers of
gravitons might be detected in colliders in a forseeable future\experiment. In
the original formulation of Randall and Sundrum, a continuous spectrum of
Kaluza-Klein modes of graviton arises without any mass gap\rs. However, to
have a well defined effective field theory, it is desirable to have a solution
where the graviton is well separated by a mass gap from the Kaluza-Klein
modes. In this paper we consider a model with the desired spectrum of
Kaluza-Klein with a mass gap without direct reference to supergravity. This is
when we have a periodic array of thin branes, in light of recent proposals for
a network such braneworlds\network,  or a folded brane\manyfold, producing
many idential braneworlds\hstt.
Such a picture raises an interesting question. What will be the effect on our
world from the presence of other worlds?
Some indirect effects such as messengers of supersymmetry
breaking\network\manyfold, or cosmological/astrophysical effects were
discussed\manyfold. Here we will discuss one another aspect of such a network
of braneworld, which might be more sensitive to the shape of the network of
the braneworlds. It is the spectrum of Kaluza-Klein modes, which might perhaps
be explorable at LC and Muon Colliders\experiment.
As is quite familiar from condensed
matter physics, which deals with periodic systems (one or higher dimensional)
in many cases, the energy levels of have a distinct feature of the periodicity.
This is the occurance of forbidden zones and allowed bands in energy
spectrum\ashcroft. If we recall that the fluctuation equations of modes around
a stable BPS configurations satisfy (supersymmetric) quantum
mechanics\cnamone, and the spectrum of Kaluza-Klein spectrum comes from the
energy spectrum of such a Schr\"odinger equation with the potential determined
by the shape of the stable BPS object, we expect that a periodic configuration
of such objects will lead to Schr\"odinger equation with a periodic potential
and the Kaluza-Klein spectrum will have mass gaps, reflecting the {\it band}
structure. One might argue that there will be no distiction between this and
the torus compactification around a circle.
However, since it is likely that the number of large extra dimensions will be
equal or larger than 2, there are possibilities of highly nontrivial
network of braneworlds, as we have witnessed from the many different ways that
nanotubes can form and affect the electronic structure\nanotube.

Here we will be mainly concerned with the Kaluza-Klein modes of the graviton.
This is because the original formulation of Randall and Sundrum necessarily
has a continuum of Kaluza-Klein modes without any mass gap, and the very low lying modes
has been a discomforting factor of the model. Here we will explicitly
calculate the mass gap arising from a periodic system of 3 branes. 
There has been some works\brandhuber\
having mass gaps from a distribution of D-branes in the context of five
dimensional supergravity, however the spectrum is different from what we
consider here.

Although we have confined our calculations to a simple one dimensional case
this method can be easily generalized to higher dimenensional cases as well as
more complicated networks, as well as other potentials for the fluctuation
equations from smooth models\bpsdom\dfgk. For example, for the graphite like
structure (or nanotube like structures) we can utilize the hexagonal symmetry
of the system and obtain the band structure, although we might have to resort
to numerical methods eventually.
Let us now consider the solution five dimensional metric that respects four
dimensional Poincar\'e invariance\rs, plus
the linearlized tensor fluctuations $h_{\mu\nu}$ around it.
\eqn\metric{ds^2 =
\left(e^{-2k|y|}\eta_{\mu\nu} +h_{\mu\nu}(x,y)\right) dx^\mu dx^\nu + dy^2.}
The spectrum $h_{\mu\nu}$ satisfies the
Schr\"odinger equation, with the appropriate change of variables\rs:
$h(x,y)=\psi(y)e^{ip\cdot x}$,
$\hat\psi(z) = \psi(y) e^{k|y|/2}, \hat h (x,z)= h(x,y) e^{k|y|/2},$ and
$z = {\rm sgn} (y) (e^{k|y|}-1)/k.$ (We will be following closely the
notations of Ref.\rs.)
One of the important feature of this model is that
the bound state of the higher dimensional graviton is localized in the extra
dimension. That is to say, the graviton is confined to a small region
within this infinite space. The argument supporting the existence goes as
follows: One considers the wave equation satisfied small gravitational
fluctuations, which takes the form of a nonrelativistic Schr\"odinger
equation:
\eqn\Schro{\left[-{1\over 2} \p^2_z +V(z) \right]\hat\psi(z) = m^2
\hat \psi(z).}
For a simple configuration of single domain wall, one has the so called
volcano-potential. \eqn\volcano{
V(z) = {15k^2 \over 8(k|z|+1)^2} - {3k \over 2} \delta (z).}
(We will be working in the units of $\hbar = 1, {\rm mass}=M=1$.)

First of all there is a single normalizable bound state mode, which is the
graviton of the 3+1 dimensional world, supported by the delta function term in
the potential:
\eqn\zeromode{\hat \psi_0 (z) = k^{-1} (k|z|+1)^{-3/2}.}
There is also a continuum Kaluza-Klein modes, with `energy' $E=m^2/2$;
\eqn\kkmodes{\hat \psi_m (z) \sim N_m (|z|+1/k)^{1/2} 
\left[Y_2(m(|z|+1/k))+ {4k^2\over \pi m^2} J_2(m(|z|+1/k))\right],}
where $N_m$ is the normalization constant.
The gravitational force in our effective four dimensional
world is due to the exchange of the zero mode as well as continuum
Kaluza-Klein modes.
These modes in the continuum without any gap potentially have some problems
phenomenologically.
To cure this, let us now consider a brane world that is a one
dimensional periodic lattice of these worlds, separated apart by a lattice
spacing of $l$. Then the potential for the gravitational fluctuation will see
a periodic potential: \eqn\periodic{{\cal V}(z) = \sum^{\infty}_{n=-\infty}
V(z-n l).}
As is familiar from the condensed matter physics, we expect that the energy
spectrum now develops a {\it band structure} with gaps between the bands.
The typical width of the gap will be related to the lattice spacing.
One of the key question for us would be the location of the first band of
allowed Kaluza-Klein modes, which will be affecting our four dimensional effective
theory most significantly.

First of all, we assume that the domain walls are
locally BPS objects. Then the fluctuation equations must be a supersymmetric
quantum mechanics, so we are guaranteed a massless mode which can be
interepreted as the graviton.
This is certainly the case for the volcano potential.
In fact we can easily check that the following
is the solution with zero mass:
\eqn\zeromodetwo{\hat \psi_0 (z) \sim\sum^{\infty}_{n=-\infty}
 k^{-1} (k|z-n l|+1)^{-3/2}.}
(We have to normalize the wave function piecewise for each period.)
So around each vocano potential, there will be a localized zero mode, and the
shape of the zero mode will be almost the same on each brane, when the branes
are well separated.

Now let us consider the Kaluza-Klein modes, whose spectrum will depend on the
detailed shape of the potential. One of the key feature of the
Kaluza-Klein modes regardless of the detailed shape will be that not all the
energy eigenvalues will be allowed.
We will be considering the cases where the
potentials are well separated, that is $l> 1/k$. We will have many problems
when the branes are overlapping with each other. In that case the wave
functions behave as sinusoidal functions for regions away form the center of
the volcano potential. This is possible because the Bessel functions $Y_n(z)$
and $J_n(z)$ behave as sinusoidal functions for large values of $z$, when
multiplied by $\sqrt{z}$. To analyze the spectrum, we first consider a problem
of a single potential. The wave function to the left of the potential is $\hat
\psi (z) \sim A e^{im z} + B e^{-im z}$ for $x\ll -1/k$ and the wave function 
to the right 
will be $\hat \psi (z) \sim C e^{im z} + D e^{-im z}$ for $x\gg 1/k$.
The relation between the coefficients will be through the transmission
matrix $M(m)$ as follows:
\eqn\tranmission{ \left(\matrix{A \cr B}  \right)
= M(m)\left(\matrix{ C \cr D}\right)
=\left(\matrix{ F(m) & G^*(m) \cr  G(m) &  F^*(m)}\right)
\left(\matrix{ C \cr D}\right).}
 In the above we have used the condition for
real potential $V(z)$ for the general form of the transmission
matrix. We have additional condition that $G(m)$ is pure imaginary if
we have even potential $V(z)= V(-z)$.

The evaluation of the components of the tranmission matrix can be done
using WKB approximation and we have the following solutions\merzbacher:
\eqn\wkb{F(m) = \theta(m) + {1\over 4\theta(m)},
\ \ \  G(m) = i \left(\theta(m) - {1\over 4\theta(m)}\right).}
where
\eqn\ttt{ \theta(m) = \exp \left(\int^b_a dz \sqrt{2(V(z)-E)}    \right).}
In the above  we have $V>E=m^2/2$ for $a<z<b$.
For the volcano potential, and for $m^2/2 < 15k^2/8$,  we have $a=
-\left({\sqrt{15}\over 2m} -{1\over k}\right)$ and $b=\left({\sqrt{15}\over
2m} -{1\over k}\right)$. (The delta function will not contribute, because
we have $V<E$.)
So for the volcano potential, we have
\eqn\tteee{{\rm ln}(\theta(m)) =
15(\log(\mu+\sqrt{\mu^2-1})-\sqrt{1-1/\mu^2}). }
where $\mu = \sqrt{15\over4} {k\over m} >1$.
The value of $\theta$ is between
one and infinity, and especially for the value of $m$ near zero, we have
$F(m)\sim \theta\sim \infty$. $F(m)$ is real throughout the values of $m$.
This large value of $F(m)$ for small $m$ originates from the fact that $V(z)$
is thick at the base of the volcano.

Now consider a potential
which is obtained by juxtaposing $N\rightarrow \infty$ of the single potential
we have considered above.
Then the condition for the allowed energy is that\cohentannouji
\eqn\condition{ X=|{\rm Re} (e^{im l} F(m) )| \leq 1.}
For the case at hand, we have
real values for $F(m)$, so that the condition \condition\ is simply
$|\cos(m l) F(m)| \leq 1.$
The function on the left hand side is oscillating with the enveloping
amplitude determined by $F(m)$. Since for all the values $F(m)\geq 1$, we
see that there are necessarily forbidden zones, because certain ranges of
values of $m$ do not satisfy the condition. Especially, for the region of $m$
near zero, $F(m)\rightarrow \infty$, so that for small enough $m$ the
condition is not satisfied, because we have  $\cos(m l) \sim 1$.
The first Kaluza-Klein mode develops where $\cos(ml)\sim 0$ for the smallest
value of $m$, since there $F(m)$ is very large. Thus we have $m\sim {\pi\over
2l}.$
We see that there is a forbidden zone right above the zero mode,
and this is a boon for the phenomenology for the following reasons.
(If we did have a phase for $F(m)$ we might not enjoy the mass gap right above
the graviton.)

 i) Since there will be a gap right above the zero mode, the
lowest Kaluza-Klein mode will start at, say, $m_k$.
For this case the physics of the effective four dimensional
theory will be mostly affected by the first band and the Newton's law of
gravity will be as follows:
\eqn\newtonlaw{ V_N(r) \sim G_N{m_1 m_2\over r} \left(1+ e^{-m_k r}
\left({m_k \over r} + {1\over r^2}\right)   \right).}
We have an exponential suppression of the correction to the Newton's law.
This will be a generic feature of any of the models involving a network of
brane worlds or `manyfold' universe, as long as there are many of the brane
worlds to have periodicity.

ii) There will be modifications in the Standard Model cross sections due to
the virtual exchange of graviton towers.
New processes not allowed in the Standard Model at the tree
level might appear.
In the case of exchange, the amplitude is proportional to
the sum over the propagators of the entire Kaluza-Klein tower and can
potentially diverge.
Usually it is dealt with brute force regularization\experiment. In the
presence of a mass gap, it will behave much better, since the lattice size of
the network of braneworlds will give a natural cutoff scale.

We conclude this paper with some general remarks.
First of all, we have considered models with branes in the thin limit.
In principle we can model the branes from (super)gravity theories, and have
smooth potentials for the fluctuations.
In considering such a model, we will necessarily encounter a class of 
SUSY quantum mechanics with periodic superpotentials, $W(x+a) = W(x)$.
SUSY quantum mechanics with periodic potentials differ from non-periodic ones
that it is possible for both isospectral potentials to support zero modes,
whereas in the nonperiodic ones either one or neither of the pair has a zero
mode\dunne.
Most general periodic potentals which can be analytically solved involve
Jacobi's elliptic functions ${\rm sn}^2(z|k)$, which in various limits become
P\"oschl-Teller potentials, which arose in the context of Kaluza-Klein spectrum alreay,
or periodic potential such as Scarf potential\scarf. It would be desirable to
have some supergravity model with lead to such a quantum mechanics system.
(Some discussions in this context can be found in Refs.\bpsdom\bs.)

Secondly, in order to have the weak scale scale, we might have to resort to
the senario of Lykken and Randall\lykken.
In this senario, one has localized graviton zero mode on the `Planck' brane
and one has another brane at distance $y_0$ away from it such that $e^{-k y_0}
= {\rm TeV}/M_{pl}$. To incorporate this idea into our frame work, we will
regard this set of two branes as our basic unit and imagine a periodic array
of them. In terms of gravity, the `Planck' branes will be dominating the
scene, and most of our analysis will be qualitatively the same. The changes in
the volcano potential will be introduced, and we can do the similar WKB
approximation to this new system and calculate the transmission matrix. We
will again get a mass gap in the Kaluza-Klein spectrum of graviton.
This is all possible as long as the lattice size is larger than the typical
size of the brane pair which  will be 5 - 10 Planck lengths.

More challenging problem would be to consider the full two or
higher dimensional array of braneworlds and consider a `condensed' universe
and Kaluza-Klein modes in such a universe.
It might be possible to have a cosmology where we have a rather chaotic
initial condition and have become condensed to a network before the
nucleo-synthesis\network\manyfold.
If there is enough symmetry, such
as hexagonal symmetry, the corresponding Kaluza-Klein spectrum can in
principle be obtained and even an experiment sensitive enough to explore the
configuration of the brane networks might be possible. Furthermore,
compactification along the large extra-dimension can affect the spectrum, just
like the different electronic structure of a nanotube has from a graphite on a
plane\nanotube.

\centerline{\bf Acknowledgements}

I have benefitted from useful conversations with K. Hori,
B. Pioline, A. Strominger and  J. Terning.
Special thanks to H. Kim and I. Park of CMT group of Harvard
on discussions of condensed matter physics.
This work is supported by Brain Korea 21 program of Korea Research Foundation
(1999).

\listrefs
\bye